\documentclass[prl,twocolumn,showpacs,preprintnumbers,amsmath,amssymb]{revtex4}
\usepackage{graphicx}
\usepackage{dcolumn,color}
\usepackage{bm}
\usepackage{epsfig}
\def\be{\begin{equation}}
\def\ee{\end{equation}}

\begin{document}
\bibliographystyle{try}
\topmargin 0.1cm
\newcounter{univ_counter}
\setcounter{univ_counter} {0} \addtocounter{univ_counter} {1}
\edef\GATCHINA{$^{\arabic{univ_counter}}$ }
\addtocounter{univ_counter}{1}
\edef\SPBUNI{$^{\arabic{univ_counter}}$ }
\addtocounter{univ_counter}{1}
\edef\HEIDELBERG{$^{\arabic{univ_counter}}$ }
\title{Casimir effect for thin films from
imperfect materials.\\
}
\author{
V.~N.~Markov ~\GATCHINA , Yu.~M.~Pis'mak~$^{2,3}$, }
\affiliation{\GATCHINA Department of Theoretical Physics, Petersburg
Nuclear Physics Institute, Gatchina 188300, Russia, }
\affiliation{\SPBUNI Department of Theoretical Physics, State
University of Saint-Petersburg, Saint-Petersburg 198504, Russia }
\affiliation{\HEIDELBERG Institute for Theoretical Physics,
University of Heidelberg, Heidelberg D-69120, Germany }
\date{\today}
\begin{abstract}
We propose an approach for investigation of interaction of thin
material films with quantum electrodynamic fields. Using main
principles of quantum electrodynamics (locality, gauge invariance,
renormalizability) we construct a single model for Casimir-like
phenomena arising near the film boundary on distances much larger
then Compton wavelength of the electron  where fluctuations of
Dirac fields are not essential. In this model the thin film is
presented by a singular background field concentrated on a
2-dimensional surface. All properties of the film material are
described by one dimensionless parameter. For two parallel plane
films we calculate the photon propagator and the Casimir force,
which appears to be dependent on film material and can be both
attractive and repulsive. We consider also an interaction of plane
film with point charge and straight line current. Here, besides
usual results of classical electrodynamics the model predicts
appearance of anomalous electric and magnetic fields.
\end{abstract}
\maketitle It is well known that the state of quantum field
theoretical system is strongly influenced by external background.
On the other hand, the fluctuations of quantum fields make
essential correction in a classical picture of interaction for
material objects. In 1948, Casimir showed that there is a
generated attraction between two parallel uncharged planes
\cite{Casimir}. Theoretical prediction of this phenomena called
the Casimir effect (CE) has been well confirmed with modern
experimental dates \cite{Moh1,Moh2, Bressi}. Nowadays, the CE
appears to be of practical importance because the quantum
phenomena of such kind can be essential for micro-mechanics and
nano-technology.

There are many theoretical results concerning the CE (see for
example recent review \cite{milton}). However, many authors being
interested in some particular aspects of the CE only, prepare
calculations in the framework of simplified models. Usually, it is
supposed that the specifics of quantum electrodynamics (QED) are
not significant, and most essential features of the CE can be
investigated in the framework of free quantum scalar  field theory
with fixed boundary conditions or $\delta$-function potentials
\cite{Jaffe2,Milton2}. By means of such methods one can obtain
quantitative description of some of the CE characteristics, but
there is no possibility of studying other phenomena generated by
interaction of the QED fields with classical background within the
same model.

In this paper we propose  a single model for investigation of all
peculiar properties of the CE for thin material films. Here the
film is presented  by a singular background (defect) concentrated
on the 2-dimensional surface. Its interaction with a photon field
appears to be completely defined by the geometry of the defect and
restrictions following from the basic principles of QED (gauge
invariance, locality, renormalizability). The locality of
interaction means that the action functional of the defect is
represented by an integral over defect surface of the Lagrangian
density which is a polynomial function of space-time point in
respect to fields and derivatives of ones. The coefficients of
this polynomial are the parameters defining defect properties. The
requirement of renormalizibility sets strong restrictions on form
of defect action in quantum field theory (QFT). The first analysis
of renormalisation problem in local QFT with defects has been made
by Symanzik in \cite{Sim}.
 Symanzik showed that in order to keep renormalizability of the bulk
QFT, one needs to add a defect action to the usual bulk action of
QFT model. The defect action must contain all possible terms with
nonnegative dimensions of parameters and not include any
parameters with negative dimensions. In case of QED the defect
action must be also gauge invariant.

     From these requirements it follows  that
for thin film (without charges and currents) which shape is
defined by equation $\Phi ( x ) = 0$, $x=(x_0,x_1,x_2,x_3)$, the
action describing its interaction with photon field $A_\mu(x)$
reads
\begin{equation}
S_{\Phi}(A)= \frac{a}{2} \int
    \varepsilon^{\lambda \mu \nu \rho}
\partial_{\lambda} \Phi ( x ) A_{\mu} ( x ) F_{\nu
\rho} ( x ) \delta ( \Phi ( x))dx \label{v1}
\end{equation}
where $F_{\nu \rho} ( x ) = \partial_{\nu} A_{\rho} -
\partial_{\rho} A_{\nu}$,
$\varepsilon^{\lambda \mu \nu \rho} $ denotes totally
antisymmetric tensor ($\varepsilon^{0123}=1$), $a$ is a constant
dimensionless parameter. Expression (\ref{v1}) is the most general
form of gauge invariant action concentrated on the defect surface
being invariant in respect to reparametrization of one and not
having any parameters with negative dimensions.

 The CE-like phenomena considered directly on the
distances  from the defect boundary much larger then Compton
wavelength of the electron are not influenced essentially by the
Dirac fields in QED because  of exponential damping of
fluctuations of those on much smaller distances ($\sim
m_e^{-1}\approx 10^{-10}cm$ for electron, $\sim m_p^{-1}\approx
10^{-13}cm$ for proton ) \cite{deffect}. Thus, by theoretical
studies of CE one can neglect the Dirac fields and use the model
of quantum electromagnetic field (photodynamic) with additional
defect action (\ref{v1}). Such models are considered in this
paper.

In order to expose some features of ones, we calculate the Casimir
force (CF) in the model for two parallel infinite plane films. We
consider also an interaction of the plane film with a parallel to
it straight line current and an interaction of film with a point
charge. For these systems we calculate electric and magnetic
fields.

 The quantitative description of all physical phenomena
caused by  interaction of the film  with the photon field and
classical charges can be obtained if the generating functional of
Greens functions is known. For gauge condition $\phi(A)=0$  it is of
the form
\begin{equation}
G(J)=C\int e^{i S(A,\Phi)+iJA}\delta({\phi(A)})DA \label{va}
\end{equation}
where
\begin{equation}
S(A,\Phi)= -\frac{1}{4}F_{\mu\nu}F^{\mu\nu}+ S_{\Phi}(A), \label{ac}
\end{equation}
and the constant $C$ is defined by normalization condition
$G(0)|_{a=0}=1$, i.e. in  pure photodynamic without defect $\ln
G(0)$ vanishes.  The first term on the right hand side of
(\ref{ac}) is the usual action of photon field. Along with defect
action it forms a quadratic in photon field full action of the
system which can be written as $ S(A,\Phi)= 1/2 \ A_\mu
K_\Phi^{\mu\nu}A_{\nu} $.
   The integral (\ref{va}) is gaussian and is
calculated exactly:
$$
G(J)=\exp \left\{\frac{1}{2}Tr \ln(D_\Phi D^{-1})-
\frac{1}{2}JD_\phi J\right\}
$$
where $D_\Phi$ is the propagator $D_\Phi=iK_\Phi^{-1}$ of
photodynamic with defect in gauge $\phi (A)=0$, and $D$ is the
propagator of photon field without defect in the same gauge. For
the static defect, function $\Phi(x)$ is time independent, and
$\ln G(0)$ defines the Casimir energy.

To take into account effects of finite width one should add to (\ref{ac}) bulk defect action
with a set of independent material parameters. As we show below bulk contribution to Casimir energy
can be neglected for thin films.

  We restrict ourselves with consideration of the simplest
case of plane infinite films. We begin with defect concentrated on
two parallel planes $x_3=0$ and $x_3=r$. For this model, it is
convenient to use a notation like $x=(x_0,x_1,x_2,x_3)=(\vec{x},
x_3)$. Defect action (\ref{v1}) has the form:
$$
 S_{2P}=\frac{1}{2}\int(a_1\delta(x_3)+a_2
\delta(x_3-\mbox{r}))
 \varepsilon^{3 \mu\nu \rho} A_{\mu}(x) F_{\nu
\rho}(x)dx.
$$
 The defect action $S_{2P}$  was discussed in (10)
in substantiation  of Chern-Simon type boundary conditions chosen
for studies of the Casimir effect in photodynamics \cite{bordag2}.
This approach based on boundary conditions is not related directly
to the one we present.  The defect action (\ref{v1}) is the main
point in our model formulation, and no any boundary conditions are
used. The action $S_{2P}$ is translationaly  invariant with respect
to coordinates $x_i$, $i=0,1,2$. The propagator $ D_\Phi(x,y)$ is
written as:
$$
D_{2P}(x,y) =\frac{1}{(2\pi)^3}\int
D_{2P}(\vec{k},x_3,y_3)e^{i\vec{k}(\vec{x}-\vec{y})}d\vec{k},
$$
and $D_{2P}(\vec{k},x_3,y_3)$ can be calculated exactly. Using latin
indexes for the components of 4-tensors with numbers $0,1,2$ and
notations
$$
P^{lm}(\vec{k})=g^{lm}- k^l k^m/{\vec k}^2, \ \
L^{lm}(\vec{k})=\epsilon^{lmn3}k_n
$$
($g$ is metrics tensor) one can present the results for the
Coulomb-like gauge $\partial_0A^0+\partial_1 A^1+\partial_2 A^2=0$
as follows
\begin{widetext}
\begin{eqnarray} D_{2P}^{lm}(\vec{k},x_3,y_3) =
\frac{ P^{lm}(\vec{k})}{2| \vec{k} |^{}} \left\{
   \frac{
   ( a_1 a_2 - a^2_1 a_2^2 ( 1 - e^{2 i|\vec{k}|
\text{r}} ) )
  ( e^{i| \vec{k} | ( |x_3 | + |y_3 - \text{r} | )} +
e^{i| \vec{k} | (
   |x_3 - \text{r} | + |y_3 | )} )  e^{i| \vec{k} |
   \text{r}}}{
   [ ( 1 + a_1 a_2 ( e^{2 i | \vec{k} | \text{r}} - 1
)
   )^2 + ( a_1 + a_2 )^2 ]}+
   \right.
   \nonumber
   \\
   \left.
   +\frac{
    ( a^2_1 + a^2_1 a_2^2 ( 1 - e^{2
   i| \vec{k} | \text{r}} ) ) e^{i | \vec{k} | ( |x_3
| + |y_3 | )} +
   ( a^2_2 + a^2_1 a_2^2 ( 1 - e^{2 i| \vec{k} |
\text{r}} )
   ) e^{i| \vec{k} | ( |x_3 - \text{r} | + |y_3 -
\text{r} | )}
    }{[ ( 1 + a_1 a_2 ( e^{2 i | \vec{k} | \text{r}} -
1 )
   )^2 + ( a_1 + a_2 )^2 ]
    } -e^{i| \vec{k} | |x_3 - y_3 |}\right\}
    -
\nonumber
\\
  - \frac{ L^{lm}(\vec{k})}{2 | \vec{k} |^2 [ ( 1 +
a_1 a_2 ( e^{2 i | \vec{k}
  |\text{r}} - 1 )
   )^2 + ( a_1 + a_2 )^2 ]} \left\{ a_1 a_2 ( a_1 +
   a_2 ) \left( e^{i| \vec{k} | ( |x_3 | + |y_3 -
\text{r} | )} + e^{i|
   \vec{k} | ( |x_3 - \text{r} | + |y_3 | )} \right)
e^{i| \vec{k} | \text{r}}
   \right.-
   \nonumber
   \\
   \left. \left. \left. - a_1 ( 1 + a_2 ( a_2 + a_1
e^{2 i| \vec{k} |
   \text{r}} ) ) \right. \right. e^{i | \vec{k} | (
|x_3 | + |y_3 | )} - a_2 (
   1 + a_1 ( a_1 + a_2 e^{2 i| \vec{k} | \text{r}} )
e^{i| \vec{k} | ( |x_3 -
   \text{r} | + |y_3 - \text{r} | )} \right\},
 \nonumber
\\
 D_{2P}^{l3}(\vec{k},x_3,y_3)=
D_{2P}^{3m}(\vec{k},x_3,y_3)=0,\ \
D_{2P}^{33}(\vec{k},x_3,y_3)=\frac{-i \delta(x_3-y_3)}{ |{\vec
k}|^2}, \ \    |\vec k|\equiv \sqrt{k^2_0-k^2_1-k_2^2} . \nonumber
\end{eqnarray}
\end{widetext}
The energy density $E_{2P}$ of defect is defined as
$$
   \ln G(0)=\frac{1}{2} \ \mbox{Tr} \ln(D_{2P} D^{-1})=-iTS E_{2P}
$$
where $ T=\int dx_0 $ is time of existence of defect, and $S=\int
dx_1 dx_2$, is area of one. It is expressed in an explicit form in
terms of polylogarithm function $\mbox{Li}_4(x)$ in the following
way:
$$
E_{2P}=\sum_{j=1}^2 E_j+E_{Cas},  E_j =
\frac{1}{2}\int\ln(1+a_j^2)\frac{d\vec{k}}{(2\pi)^{3}} , j=1,2,
$$
\begin{eqnarray}
E_{Cas}=-\frac{1}{16 \pi^2
\mbox{r}^3}\sum_{k=1}^2\mbox{Li}_4\left(\frac{a_1a_2}{a_1a_2+i(-1)^k
(a_1+a_2)-1}\right) \nonumber
\end{eqnarray}
Here $E_j$ is an infinite constant, which can be interpreted as
self-energy density on the $j$-th planes, and $E_{Cas}$ is an energy
density of their interaction.
 Function $\mbox{Li}_4(x)$ is defined as
$$
\mbox{Li}_4(x)=\sum_{k=1}^\infty \frac{x^k}{k^4}=-\frac{1}{2}\int_0
^\infty k^2 \ln(1-xe^{-k})dk.
$$
For identical defect planes ($a_1=a_2=a$) the force
$F_{2P}(\mbox{r},a)$ between them is given by
$$
F_{2P}(\mbox{r},a)=- \frac{\partial E_{Cas}(\mbox{r},a)}{\partial
\mbox{r}}=- \frac{\pi^2}{240 \mbox{r}^4} f(a).
$$
Function $f(a)$ is plotted on the Fig. $1$. It is even
($f(a)=f(-a)$) and has a minimum at $|a|=a_m \approx 0.5892$
($f(a_{m})\approx 0,11723$), $f(0)=f(a_0)=0$ by $a_0\approx
1,03246$, and
 $\lim_{a\rightarrow \infty}f(a)=1$.
For $ 0<a< a_0$ ($a>a_0$), function $f(a)$ is negative (positive).
Therefore the force $F_{2P}$ is repulsive for $|a|< a_0$ and
attractive for $|a|>a_0$. For large $|a|$ it is the same as the
usual CF between perfectly conducting planes. This model predicts
that the maximal magnitude of the {\it repulsive} $F_{2P}$ (about
$0,1$ of the CF's magnitude  for perfectly conducting planes) is
expected for $|a|\approx 0.6 $.
\begin{figure}
\centerline{\epsfig{file=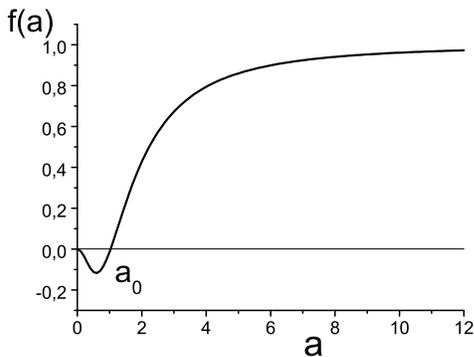,width=7cm}} \caption{Function
$f(a)$ determining Casimir force between parallel planes}
\end{figure}
For two infinitely thick parallel slabs the repulsive CF was
predicted  also in \cite{kenneth}.

 In real physical situations the film has a finite width, and the
bulk contributions to the CF for nonperfectly conducting slabs with
widths $h_1$, $h_2$ are proportional to $h_1 h_2$. Therefore it
follows directly from the dimensional analysis that the bulk
correction $F_{bulk}$ to the CF is of the form $F_{bulk} \approx
cF_{Cas}h_1h_2/r^2$ where $F_{Cas}$ is the CF for perfectly
conducting planes and $c$ is a dimensionless constant. This
estimation can be relevant for modern experiments on the CE. For
instance, in \cite{Bressi} there were results obtained for parallel
metallic surfaces where width of layer was about $h \approx 50$ nm
and typical distance $r$ between surfaces was $0.5 \mu m \leq r \leq
3 \mu m$. In that case $3\times 10^{-4}\leq (h/r)^2 \leq 10^{-2}$.
In \cite{Bressi} authors have fitted the CF between chromium films
with  function $C_{Cas}/r^4$. They claim that the value of $C_{Cas}$
coincides with known Casimir result within a $15\%$ accuracy.  It
means that bulk force can be neglected, and only surface effects are
essential. In our model the values $a>4.8$ of defect coupling
parameter $a$ are in good agreement with results of \cite{Bressi}.

To reveal some specific effects generated by interaction of the
film with the photon field within the proposed model, one can
study the scattering of electromagnetic waves on the plane and
coupling of plane with a given classical 4-current.

For the first
problem one can use a  homogeneous classical equation
$K^{\mu\nu}_{2P} A_\nu=0$ of simplified model with $a_1=a$,
$a_2=0$. It has a solution in the form of a plane wave. If one
defines transmission (reflection) coefficient as a ratio of
transmitted  $U_t$  (reflected $U_r$) energy to  incident $U_{in}$
one: $ K_t = U_t/U_{in}$, ($K_r = U_r/U_{in}$), then direct
calculations give the following result: $ K_t = (1 + a^2 )^{-1},
K_r = a^2(1 + a^2)^{-1} $. We note two features of reflection and
transmission coefficients. The first one is that in the limit of
infinitely large defect coupling these coefficients coincide with
coefficients for a perfectly conducting plane. The second one is
that they do not depend on the incidence angle, and that can be
attributed to $\epsilon \mu =1$ condition discussed below.

The classical charge and the wire with current near defect plane are
modeled by appropriately chosen 4-current $J$ in (\ref{va}). The
mean vector potential $ {\cal A}_\mu $ generated by $J$ and the
plane $x_3=0$, with $a_1=a$ can be calculated as
\begin{equation}
 {\cal
A}^\mu=-i\frac{\delta G(J)}{\delta J_\mu}\Bigg|_{a_1=a,a_2=0}=
iD_{2P}^{\mu\nu}J_{\nu}|_{a_1=a,a_2=0}. \label{vecpot}
\end{equation}
Using notations ${\cal F}_{ik}=\partial_i {\cal A}_k-\partial_k
{\cal A}_i$, one can present electric and magnetic fields as ${\vec
E}=({\cal F}_{01},{\cal F}_{02},{\cal F}_{03})$, ${\vec H}=({\cal
F}_{23},{\cal F}_{31},{\cal F}_{12})$. For charge $e$ at the point
$(x_1,x_2,x_3)=(0,0,l)$, $l>0$ the corresponding classical 4-current
is
$$
J_{\mu}(x)=4  \pi e \delta{(x_1)} \delta{(x_2)}
\delta{(x_3-l)}\delta_{0\mu}
$$
In virtue of (\ref{vecpot}), the electric field in considered
system is the same as one generated in usual classical
electrostatic by charge $e$ placed on distance $l$ from infinitely
thick slab with dielectric constant $\epsilon=2a^2+1$. The defect
plane induces also a magnetic field ${\vec H}=(H_1, H_2, H_3)$:
$$
H_1=\frac{e a x_1}{(a^2+1)\rho^3}, \ H_2=\frac{e a
x_2}{(a^2+1)\rho^3},\ H_3=\frac{e a(|x_3|+l)}{(a^2+1)\rho^3}.
$$
where $\rho=(x_1^2+x_2^2+(|x_3|+ l)^2)^{\frac 12}$. It is an
anomalous field which doesn't arise in classical electrostatics.
Its direction depends on sign of $a$. A current with density $j$
flowing in the wire along the $x_1$-axis is modeled by
$$
J_{\mu}(x)=4 \pi j  \delta(x_3-l) \delta(x_2) \delta_{\mu 1}
\label{vb10}
$$
For magnetic field from (\ref{vecpot}) we obtain usual results of
classical electrodynamics for the current parallel to infinitely
thick slab with permeability $\mu=(2a^2+1)^{-1}$. There is also an
anomalous electric field $\vec{E}=(0, E_2, E_3)$:
$$
E_2=\frac{2j a}{a^2+1} \frac{x_2}{\tau^2}, \ E_3=\frac{2j a}{a^2+1}
\frac{|x_3|+l}{\tau^2}
$$
where $\tau=(x_2^2+(|x_3|+ l)^2)^{\frac 12}$. Comparing both
formulae for parameter $a$ we obtain the relation $\epsilon \,
\mu=1$. It holds for material of thick slab  interaction of which with
point charge and current in classical electrodynamics was compared
with results for thin film of our model. The speed of light in this
hypothetical  material is equal to one in the vacuum. From the
physical point of view, it could be expected, because interaction of
film with photon field is a surface effect which can not generate
the bulk phenomena like decreasing  the speed of light in the
considered slab.

   The relation $\epsilon \, \mu=1$ is not new in the context of the
Casimir theory. It was first introduced by Brevik and Kolbenstvedt
\cite{Brevik82} who calculated the Casimir surface force density
on the sphere. Only on this condition a contact term turn out to
be zero \cite{Brevik82}. It has been investigated in a number of
subsequent papers. In our approach this condition arises naturally
because we have only one parameter $a$ that must describe both
magnetostatic and electrostatic properties of the film. It is
quite possible that macroscopic properties of surface are
different as bulk ones due to the loss of periodicity of lattice
potential in one dimension. This happens because the presence of
surface leads to a change of spectrum of electron states on and
near the surface \cite{prutton}. The essential property of
interaction of films with classical charge and current is the
appearance of anomalous fields. This fields are suppressed in
respect of usual ones by factor $a^{-1}$ and they vanish in case
of perfectly conducting plane. Magnetoelectric (ME) two-dimensional
materials \cite{Novoselov} are good candidates to detect anomalous
fields and non ideal CE. It is important to note that  for ME
films the Lifhitz theory of Casimir effect is not relevant but
they can be studied in our approach.

   The main result of our study on the CE for
films in the QED is the following. We have shown that if the CF
holds true for thin film from imperfect material, then an
interaction of this film with the QED fields can be modeled by
photodynamic with the defect action (\ref{v1}) obtained by most
general assumptions consistent with locality, gauge invariance and
renormalizability  of model. Thus, basic principles of QED were
essential in our studies of the CE. These principles make it
possible to expose new peculiarities of the physics of macroscopic
objects in QED and must be taken into account for construction of
the models. For plane films we have demonstrated that the CF is not
universal and depends on properties of the material represented by
the parameter $a$. For $a\rightarrow\infty$ one can obtain the  CF
for ideal conducting planes. In this case the model coincides with
photodynamic considered in \cite{bordag} with boundary condition
$\epsilon^{ijk3}F_{jk}=0$ ($i=0,1,2$) on orthogonal to the
$x_3$-axis planes. For sufficiently small $a$ the CF  appears to be
repulsive. Interaction of plane films with charges and currents
generate anomalous magnetic and electric fields which do not arise
in  classical electrodynamics. The ME materials could be used  for
observation of  phenomena predicted by our model.
 We hope that the obtained theoretical results can be proven
by modern experimental methods.

\section*{Acknowlegements}
We thank  D.V.Vassilevich and M. Bordag for valuable discussions.
The work  was supported  by Grant 05-02-17477 (V.N. Markov) and
Grant 03-01-00837 (Yu.M. Pis'mak) from Russian Foundation for Basic
Research.

\end{document}